\documentclass{PoS}
\usepackage{natbib}
\usepackage{graphicx}
\usepackage{sidecap}
\title{Astronomy below the Survey Threshold}

\ShortTitle{Sub-threshold astronomy in the SKA era}

\author{
\speaker{Jonathan T.~L.~Zwart}$^1$, 
{Jasper Wall}$^2$,
{Alexander Karim}$^3$,
{Carole Jackson}$^4$,
{Ray Norris}$^5$,
{Jim Condon}$^6$,
{Jose Afonso}$^{7,8}$,
{Ian Heywood}$^5$,
{Matt Jarvis}$^{1,9}$,
{Felipe Navarrete}$^{3}$,
{Isabella Prandoni}$^{10}$,
{Emma Rigby}$^{11}$,
{Huub Rottgering}$^{11}$,
{Mario Santos}$^1$,
{Mark Sargent}$^{12}$,
{Nick Seymour}$^4$,
{Russ Taylor}$^{1,13}$,
{Tessa Vernstrom}$^2$
\\ 
$^1$University of the Western Cape, Department of Physics, Private Bag
X17, Bellville 7035, South Africa\\
$^2$University of British Columbia, Department of Physics and
Astronomy, 6224 Agricultural Road, Vancouver, V6T\,1Z1, Canada\\
$^3$Max-Planck-Institut f\"ur Radioastronomie, Auf dem H\"ugel 69, D-53121 Bonn, Germany
\\
$^4$International Centre for Radio Astronomy Research, Curtin
University, Bentley, WA 6102, Australia\\
$^5$CSIRO Astronomy \& Space Science, PO Box 76, Epping NSW 1710,
Australia\\
$^6$National Radio Astronomy Observatory, 520 Edgemont Road,
Charlottesville, VA 22903, USA\\
$^7$Instituto de Astrof\'{i}sica e Ci\^{e}ncias do Espa\c co, Universidade de Lisboa, OAL, Tapada da Ajuda, PT1349-018 Lisboa, Portugal\\
$^8$Departamento de F\'{i}sica, Faculdade de Ci\^{e}ncias,
Universidade de Lisboa, Edif\'{i}cio C8, Campo Grande, PT1749-016
Lisbon, Portugal\\
$^9$Astrophysics, Denys Wilkinson Building, Keble Road, University of
Oxford, Oxford, Oxon, OX1 3RH, UK\\
$^{10}$INAF, Istituto di Radioastronomia, Via P Gobetti 101, 40129 Bologna, Italy\\
$^{11}$Leiden Observatory, Leiden University, P.O.~Box 9513, 2300 RA
Leiden, The Netherlands\\
$^{12}$Astronomy Centre, Dept.~of Physics and Astronomy, School of Maths and Physical Sciences,
University of Sussex, Falmer, Brighton BN1 9QH, UK\\
$^{13}$Astrophysics, Cosmology \& Gravity Centre, Department of Astronomy, 
University of Cape Town, Private Bag X3, Rondebosch 7701, South Africa\\
E-mail: \email{jz@uwcastro.org}
}
 
\FullConference{
Advancing Astrophysics with the Square Kilometre Array\\
June 8-13, 2014\\
Giardini Naxos, Sicily, Italy}

\abstract{Astronomy at or below the {\it survey threshold}
  has expanded significantly since the publication of the original
  {\it Science with the Square Kilometer Array} in 1999 and its update in 2004. 
The techniques in this regime may be
  broadly (but far from exclusively) defined as {\it confusion} or {\it P(D)} analyses
  (analyses of one-point statistics), and {\it stacking},
  accounting for the flux-density distribution of noise-limited images
  co-added at the positions of objects detected/isolated in a
  different waveband. Here we discuss the relevant issues, present
  some examples of recent analyses, and consider some of the
  consequences for the design and use of surveys with the SKA and its pathfinders.}

\newcommand{\skipthis}[1]{}

\begin{document}

\section{Astronomy beneath the Survey Threshold}

\noindent
Since the publication of the original SKA science case
\citep{scicase1999}, and indeed since the 2004 publication of its
update by \citeauthor{car04}, more has been understood about
sub-threshold astronomy and numerous detailed analyses have shown how
to extract astronomical and cosmological results. We look to summarize
these developments, issues and techniques, as well as to note some
implications for the SKA, its precursors and pathfinders. We largely
restrict our discussion to the radio continuum, while noting that
sub-threshold techniques are highly applicable (a) to other wavebands
(e.g.~\citealt{blast2009}, \citealt{bourne2012}), (b) when using
polarization to investigate cosmic magnetism (\citealt{chapter-stil}),
and (c) for HI spectral-line work (\citealt{chapter-blyth}).

Source surface densities generally rise steeply with decreasing source
intensity --- i.e.~the {\it source count} is `steep'. With this come
two aspects of surveys now generally accepted.  Firstly, it is very
unwise to `push' cataloguing and source-counting --- the
\textit{survey threshold} --- down to anywhere near the {\it confusion
  limit} or the {\it noise limit}; the combined effects of these plus
{\it Eddington bias} require cataloguing to be limited to intensities
> 5$\sigma$, $\sigma$ representing the combined confusion (see below)
and system noise. Second, there is much astronomy to be done with
proper and rigorous statistical interpretation of survey data to
levels well below this threshold.

Confusion analysis ({\it P(D)}, or single-point statistics) as first
proposed by \cite{sch57}
is essentially
single-frequency, extending our knowledge of the source count at that
frequency to intensities below the threshold.
Stacking on the other hand was a conscious product of multi-waveband
astronomy, to investigate bulk properties of objects above thresholds
and in catalogues in one waveband, but generally (not exclusively)
below survey thresholds in another band.  Stacking then implies adding
images in the latter band at the positions of the catalogue in the
former, to discover average faint intensities. It was perhaps
pioneered in the X-ray regime (\citealt{caillault1985}) for the
average X-ray properties of G-stars. By the time of the SDSS survey,
sophisticated analyses were carried out to examine radio properties of
radio-quiet quasars \citep{white2007}, and indeed radio-quiet galaxies
\citep{hodge2008}. Chief requirements are absolute astrometry to
better than 1 arcsec in both bands and good dynamic range in the
sub-threshold band.

It is essential for galaxy-formation and evolution studies to be able
to investigate both faint AGN and star-formation activity (see
e.g.~\citealt{smolcic2008,smolcic2009a,smolcic2009b,seymour2009,bonzini2013,padovani2014}). This
is particularly so in view of the now commonly accepted link between
an AGN phase and the quenching of star formation
(\citealt{begelman2004,croton2006}) in order to effect cosmic
downsizing. In addition, the statistical techniques discussed here
bring within reach a wealth of extragalactic astrophysics considered
further in
\cite{chapter-jarvis,chapter-kapinska,chapter-kim,chapter-murphy,chapter-smolcic}. These
make it possible to proceed from noisy flux measurements to source
counts, luminosity functions, star-formation rates and their cosmic
densities, all as a function of e.g.~environment, stellar mass,
redshift, but {\it only} when ancillary data are available. More
comprehensive introductions to the science available via stacking and
$P(D)$ analyses (together with comprehensive referencing) are provided
in \citet{gle10, padovani2011,hey13,zwa14}.

\section{Beating the Survey Threshold}
\label{sec:defns}

\begin{figure}[b]
\includegraphics[angle=-90,trim=32mm 20mm 60mm 20mm,clip,width =0.95\linewidth]{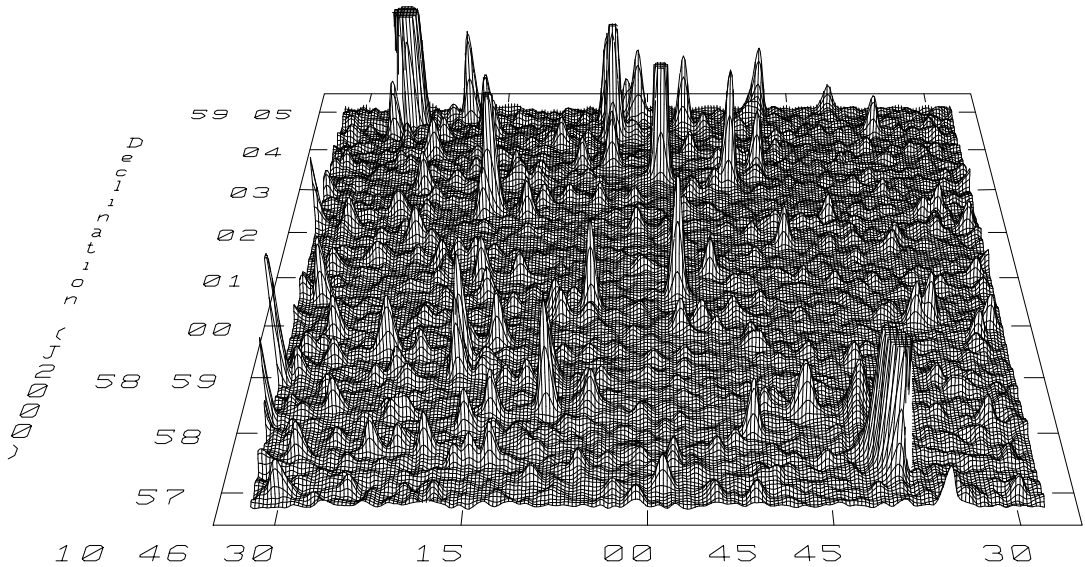}
\caption[]{A 3-GHz image from a 57-hour integration with the Karl
  G. Jansky Very Large Array (VLA) in C-configuration \protect\citep{con12},
  with synthesized beam 8-arcsec FWHM. The instrumental noise is
  1.02~$\mu$Jy beam$^{-1}$. All features in the image, positive and
  negative, are on beamwidth scales and these cover the image,
  indicating that the background of faint sources has been completely
  smoothed or integrated by the beam response.
  The large positive deflections represent discrete radio sources,
  producing the long positive tail of the $P(D)$ distribution
  (\protect\citealt{con12,ver14}; see Fig.~\protect\ref{fig2}).}
\label{fig1}
\end{figure}

\begin{figure}
\includegraphics[trim=0 12 0 0,clip,width=0.45\linewidth]{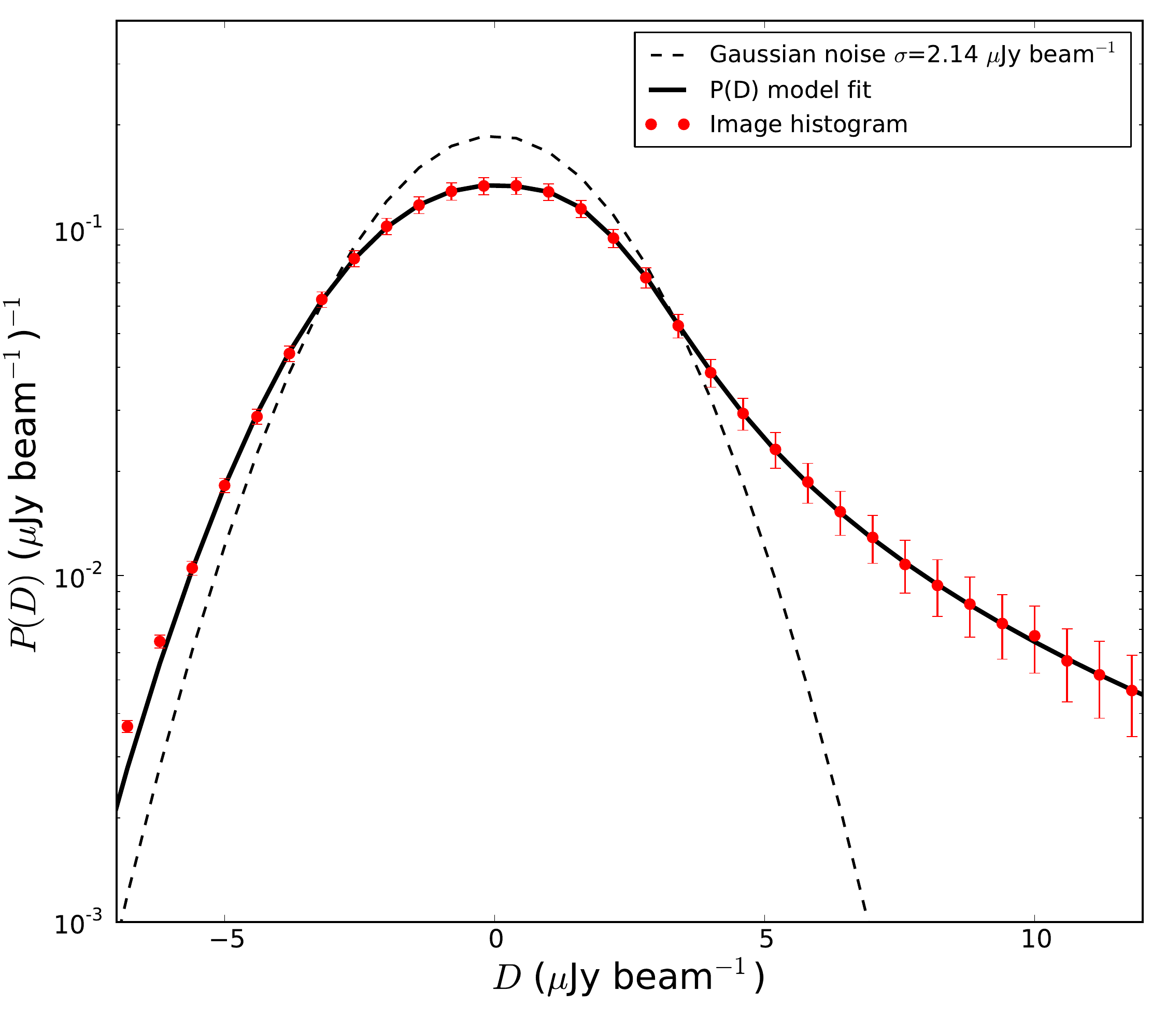}
\hfill
\includegraphics[trim=0 0 0 0 0,clip,width =0.50\linewidth,height=6cm]{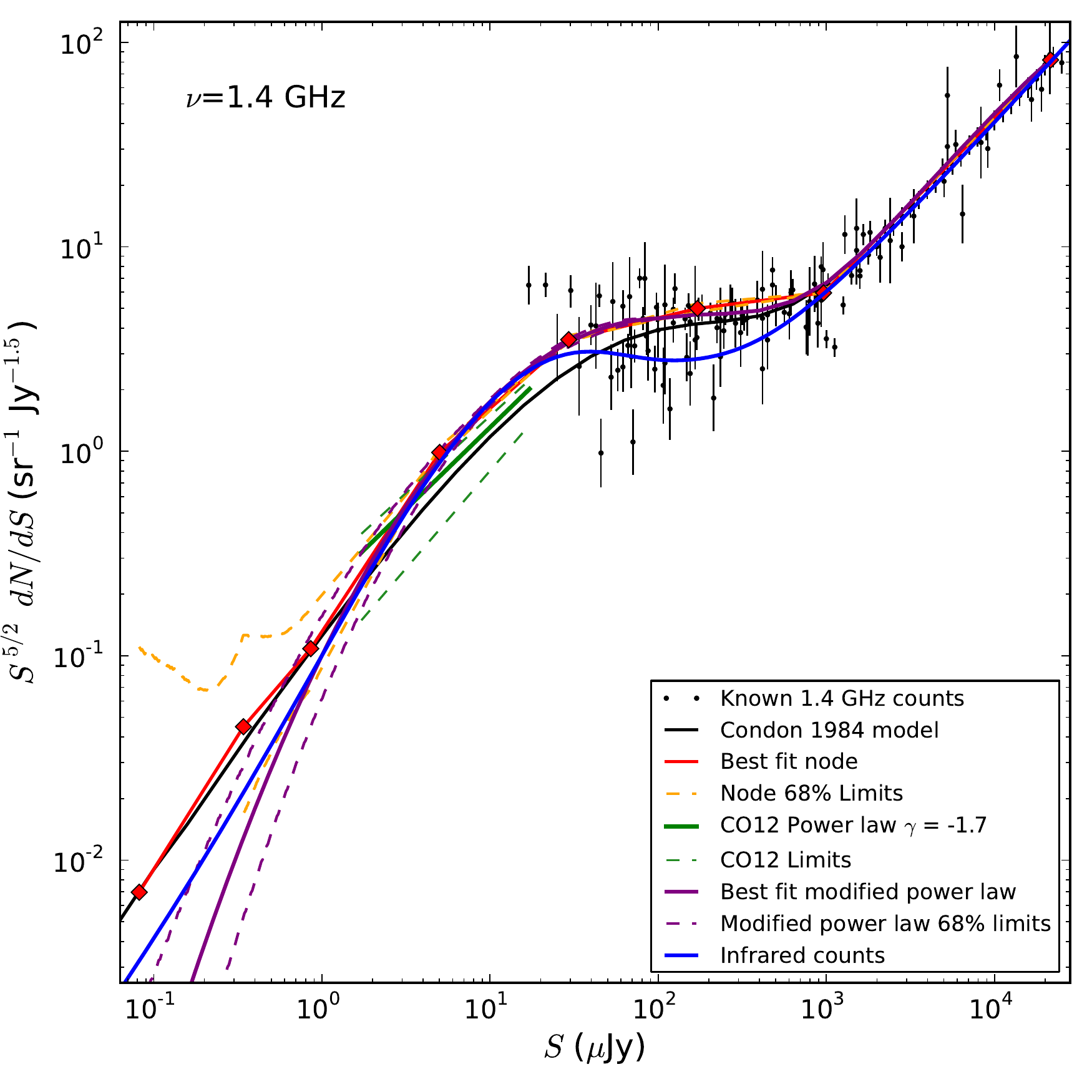}
\caption[]{{\bf Left} --- Oversampled {\it P(D)} distribution
  \protect\citep{ver14} for the central 5-arcmin radius of the 3-GHz image of
  Fig.~\ref{fig1} (dots and $\sqrt{N}$ error bars), with the
  best-fitting 8-node model (solid curve). The dashed line represents
  a Gaussian of $\sigma=$1.255~$\mu$Jy, accurately describing the
  instrumental noise.  {\bf Right} --- Source counts at 1.4 GHz, shown
  in relative differential form.  Points and error bars show previous
  estimates of the counts from the literature. The short straight
  (green) line plus dashed lines above and below it show results from
  \protect\cite{con12}, the early interpretation of the current deep survey
  data. Joined points plus dashed error limits show the results from
  the $P(D)$ analysis of \protect\cite{ver14}, translated from 3.0 to
  1.4\,GHz with a spectral index of $-$0.7. Smooth curves represent
  counts calculated from models of radio and infra-red luminosity
  functions and cosmic evolution; see \protect\cite{ver14} for references.}
\label{fig2}
\end{figure}

\noindent
Before describing the methods of $P(D)$ and stacking --- and their
biases --- in more detail, we set out some relevant
definitions. Confusion and associated concepts are discussed in the
following points, and these may be clearer on examination of
Figs~\ref{fig1} and \ref{fig2}.

\begin{figure}
\includegraphics [trim=0 0 0 0]{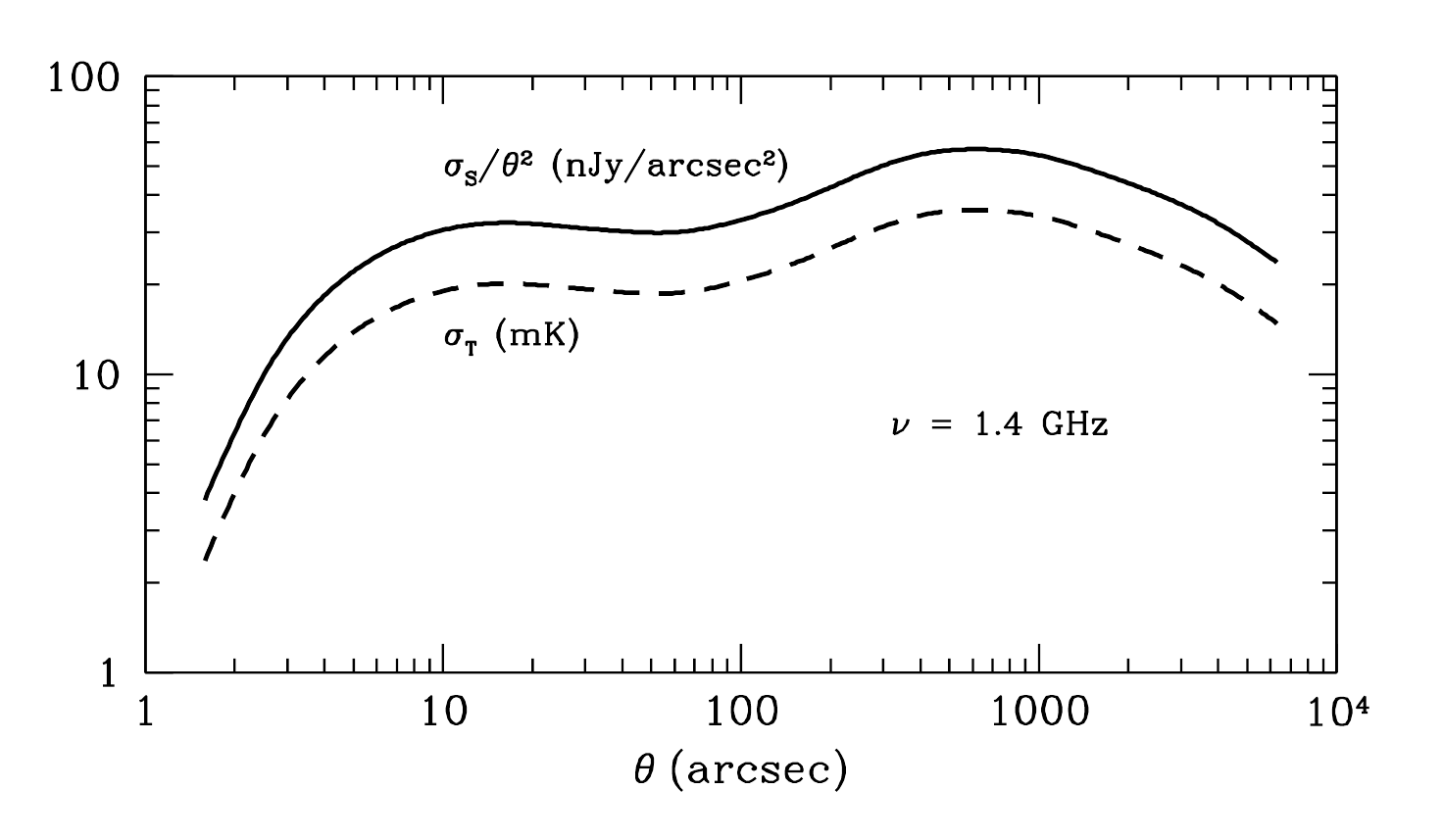}
\caption[]{1.4-GHz `rms' confusion $\sigma$ for point sources based on
  the \protect\cite{wil08} source count, as a function of FWHM resolution for
  a Gaussian beam in units of Jy\,beam$^{-1}$ and K of brightness
  temperature. The `rms' is defined in this diagram as half the
  Jy\,beam$^{-1}$ or K range containing two thirds of the data points,
  making it a more stable statistic given the long tails of the $P(D)$
  distribution. The curves can be rescaled to nearby frequencies by
  scaling $\sigma_{\mathrm{S}}\propto \nu^{-0.7}$ or
  $\sigma_{\mathrm{T}}\propto\nu^{-2.7}$.}
\label{fig:condon}
\end{figure}

\begin{itemize}
\item[]{\textbf{Confusion}} is integration or blending of the
  background of faint sources by either (a) the finite synthesized
  beam (spatial response) of a telescope, or (b) the intrinsic angular
  extents of the sources. The anisotropies of blended background
  sources become visible when instrumental noise is low enough.

\item[]{\textbf{Intrinsic (Natural) Confusion}}, also sometimes
  referred to as \textbf{Source Blending} (which has still other meanings),
  occurs if extended sources/objects in \textit{any} image --- at any
  wavelength --- physically overlap on the sky. Note that this is
  independent of the resolution of the image, as it depends only on
  the intrinsic (distribution, angular-diameter and density)
  properties of the sources. For 1.4-GHz radio surveys, intrinsic
  confusion occurs at well below 1-$\mu$Jy rms \citep{wind2003}.

\item[]{\textbf{Instrumental Confusion}} occurs if the source density
  is so high that sources are likely to be detected in a `significant'
  fraction of the resolution elements (PSFs/synthezised beams) in an
  image. 
  This is the most commonly assumed form of confusion and it generally
  comes to mind when the term is used.

\item[]{\textbf{Sidelobe Confusion}} happens if bright sources are
  likely to appear in the sidelobes of the synthesized beam to the
  extent that the \textbf{Dynamic Range} of the image is compromised
  (\citealt{chapter-sphe}; Smirnov et~al.~in prep.).

\item[]{\textbf{Identification Confusion}} occurs when the combination
  of resolution and positional accuracy in the radio image and the
  density of sources at a cross-identification wavelength
  (e.g.~optical or infra-red), are such that a `significant' fraction
  of the sources cannot be reliably identified with objects seen at
  the other wavelength. The likelihood ratio is an example of a method
  that can be used to overcome this (see e.g.~\citealt{mcalpine2012}).

\item[]{\textbf{Confusion Limit}} There are various meanings. If used
  in the context of surveys for discrete sources, the term is
  sometimes used to signify the flux-density depth to which discrete
  sources can be reliably detected and their intensities measured. In
  the foregoing we have referred to this as the \textit{survey
    threshold}. At the brighter end of the source count with steep
  slope of about $-$2.7, a safe threshold is the flux density at which
  the surface density is perhaps 50 beam-areas per source; at faint flux
  densities where the count slope is less steep, say $-$1.7, it
  becomes $\approx$25 beam-areas per source \citep{condon-report}. The
  confusion limit, the level at which the background is totally
  blended as depicted in Fig.~\ref{fig1}, must occur at flux densities
  corresponding to source densities of $>$ one source per beam area.
  The confusion limit is also taken at times to mean the flux density
  at which the {\it confusion noise} is equal to system noise.

\item[]{\textbf{Confusion Noise}} is generally taken to be a
  single-point statistic describing the width of the distribution of
  single pixel values in a confusion-dominated image; see
  Figs~\ref{fig1} and \ref{fig2}. The origin of this `noise' depends
  on which type of confusion is in question. Since the single-point
  distribution is usually highly skewed and very non-Gaussian
  \citep{condon-report}, a single descriptor of it is not suitable.
  Theoretically the distribution for a power-law source count has an
  infinite tail to the positive side so that the mean and variance are
  infinite. (With very `steep' counts approaching $-$3 in slope, the
  distribution does approximate Gaussian.) `Shallow' slopes result in
  a preponderance of brighter source intensities and a long positive
  tail. Despite the dangers, some approximation to the core width is
  generally made to compare with system noise or in rough calculations
  of attainable survey depth. Under some circumstances
  (e.g.~\citealt{mcadam}) an estimate of the confusion noise per beam
  is taken to be the second moment of the residual differential source
  count up to some limiting threshold, where all sources with flux
  densities greater than that threshold have already been subtracted;
  one feature of/problem with this definition is that confusion
  noise is then a sensitive function of survey noise i.e.~integration
  time. Fig.~\ref{fig:condon} shows a way to scale the instrumental
  confusion to different resolutions and/or frequencies near 1.4~GHz
  (see also \citealt{con12}).
 
\item[]{\textbf{Eddington bias}} is the apparent steepening of the
  observed source count \citep{eddington1913} by the
  intensity-dependent over-estimation of intensities, due to either
  system noise or confusion noise or both. The process of
  over-estimating flux densities of faint sources is sometimes called
  flux boosting, a misleading term because no physical increase of
  observed intensity is taking place, as it is in the case of
  relativistic beaming. \citet{jauncey1968} first drew `flux
  boosting' to our attention; it was the submm astronomers who
  identified it as leading to Eddington bias and grasped how to
  estimate unbiased flux densities (and hence counts) using a Bayesian
  Likelihood Analysis (BLA) and count estimates as priors
  (\citealt{wall2003,coppin2005}).

\item[]{\textbf{Estimators and biases}} Flux densities measured by
  standard techniques, e.g.~PSF fitting, or average pixel value near a
  map peak, are {\it not} flux densities, but like all measured
  quantities, are {\it estimators} of flux densities. If instrumental or
  confusion noise is significant and the source count strongly favours
  faint intensities, the usual case, then such an estimator will be
  {\it biased}. The flux density is not boosted --- the measurement is
  wrong. De-biasing, deboosting or however it is termed, is simply
  stating that a better --- possibly even unbiased --- estimator is
  being used.

\item[]{\textbf{Probability of deflection}, \textbf{P(D)}}, or
  {\textbf{P(D) distribution}}, is the full distribution of
  single-pixel values. {\textbf{P(D) analysis}} is the analysis of
  this distribution (see Section \ref{sec:pofd}) to deduce the
  underlying faint source count.

\item[]{\textbf{Stacking}} (Section \ref{sec:stacking}) is the
  co-addition of maps at the positions of sources detected in another
  map or catalogue.
  In contrast to a $P(D)$ blind analysis, stacking is intrinsically
  prior-based source extraction, but can also be used to conquer
  confusion.

\end{itemize}

\noindent
We now discuss strategies for analysing noise-limited data, be they
confusion-limited or system-noise limited. The two prevailing
techniques, which are complementary, are either blind ($P(D)$
analysis) or prior-based (stacking).

\subsection{Probability of Deflection (\textit{P(D)}) analysis}
\label{sec:pofd}

\noindent
The technique was first applied by \citet{hew61} to estimate counts at
178~MHz by forward modelling, using the 4C survey data.
{\it P(D)} has been adopted to estimate source counts (with subsequent
constraints on luminosity functions) at a number of wavebands other
than radio, e.g.~X-ray \citep{tof92,bar94}; infra-red \citep{jen91,oli92};
far infra-red \citep{fri02}; and submm \citep{pat09}. Most recently \citep{ver14} it has been
applied to the confusion-limited data of the VLA single-pointing \citet{con12} 
integration (Fig.~\ref{fig1}). A Bayesian Likelihood Analysis was used, with a
multi-node count model \citep{pat09}.
The results are in Fig.~\ref{fig2}, which
demonstrates how the {\it P(D)} analysis has resolved the discrepant
direct counts and has extended knowledge of the 3-GHz count to nanoJy
levels.

$P(D)$ analysis in the face of extended source structures is
highly problematic.
P(D) analyses to date have assumed no effects of resolution,
i.e.~assume source angular sizes << beam sizes. It is possible to
attempt such analyses on the basis of (risky) asumptions about angular
structures. A different approach was adopted by \cite{vernstrom2014}.
Their analysis of ATCA data at 1.75~GHz obtained with a $\approx$
1-arcmin beam suggested an excess of diffuse emission on scales of
1--2 arcmin.
To carry out a correct $P(D)$ analysis in the face of source
resolution generally requires detailed knowledge of the angular-size
distribution of sources as a function of faint flux density. For
example, SKA1 surveys could be used to deduce the angular-size
distribution, which could then be extrapolated to fainter levels to
aid the $P(D)$ analysis. But any population for which we had such
detailed knowledge would probably not require data from $P(D)$
analyses. Moreover any tacit assumption that we `know' the composition
of the faint-source populations to be star-forming, and that therefore
we can infer faint radio-source structures from optical data, is
premature.

\subsection{Stacking}
\label{sec:stacking}

\noindent
The literature is very ambiguous on the precise definition of
stacking, perhaps the most succinct being ``taking the covariance of a
map with a catalogue'' \citep{blast2009}. Having made some selection of
sources in an often-deeper (in terms of raw surface density)
catalogue, one measures the flux in a map, usually at another
wavelength, building up a distribution of map-extracted fluxes for
that sample. For example, one might select sources in the near
infra-red (i.e.~by stellar mass;
e.g.~\citealt{dunne2009,karim2011,zwa14}) and measure their 1.4-GHz
fluxes, using photometric redshifts to convert those
fluxes into star-formation rates via calibration to the
far-infra-red--radio correlation (e.g.~\citealt{yun2001,condon2002}).
Or one might stack a total-intensity catalogue in polarization in
order to investigate cosmic magnetism \citep{chapter-stil}. The flux
histogram thus obtained resembles the $P(D)$ of Fig.~\ref{fig2}
(left-hand dotted line), and can be binned by some physical quantity,
e.g.~stellar mass and/or redshift.

Sufficiently rich surveys (with respect to areal coverage/depth)
to-date routinely beat individual detection/confusion limits by an
order of magnitude to recover the average (see below) properties of
pre-selected source populations.
The main complications arise if the angular resolution of the map to
be stacked is much coarser than in the selection band. For
coarse-resolution maps, signal from more than one source can
contribute to the net flux in a given pixel in the stacking band
(e.g.~\citealt{webb2004}, \citealt{greve2010}) and due consideration
must be given to this as a possible source of confusion noise.
However, similar problems can in principle arise even at high angular
resolution as the source density increases for deeper maps. One might
hence conclude that a stacking experiment suffers under
low-angular-resolution conditions as well as when survey sensitivity
is increasing.
The effects of source crowding are negligible if the source
distribution is random, following a Poisson distribution at least at
the scale of the beam (\citealt{blast2009}; Viero et al. 2012). Even
if the latter condition were violated, efficient deblending algorithms
have been proposed (e.g.~\citealt{kg2010}).

Which average? A key difficulty (see e.g.~\citealt{white2007}) is what
summary statistic to use to describe the flux distribution (as a whole
or within each bin), if any. Since the brightest discrete sources
generate a long tail to higher fluxes, the median is generally
preferred over the mean. But there are a number of biases that can
corrupt this estimator, including the flux limits and shape of the
underlying intrinsic (to the population) distribution, as well as the
magnitude of the map noise \citep{bourne2012}. Indeed, much current
work is focussed on the identification and elimination of sources of
bias, and even circumventing entirely the difficult need for debiasing
a single summary statistic.

One promising avenue is the development of algorithms for modelling
the underlying source-count distribution (or other/intrinsic physical
property) parametrically in the presence of measurement noise (see
e.g.~\citealt{con13}), without the use of a single (usually biased)
summary statistic. \citet{ketron2013} demonstrated such a technique on
a COSMOS catalogue and FIRST maps, recovering the COSMOS source counts
correctly from the FIRST data using a Maximum-Likelihood method to
reach $0.75\sigma$ with 500 sources.

Zwart et al.~(in prep.) have extended this work to a fully-Bayesian
framework, allowing model selection through the Bayesian evidence, to
measure 1.4-GHz source counts for a near-infrared-selected
sample. Early indications are that the survey threshold can be beaten
by up to two orders of magnitude. \citet{ros14} also used a BLA to
measure luminosity functions, as a function of redshift, right from
stacked fluxes. And Johnston, Smith and Zwart (in prep.)  are
incorporating redshift evolution directly into the modelling process.
We also note another stacking technique
(\citealt{lindroos2013,lindroos2014,chapter-knudsen}) in which
visibilities are calculated at the positions of known sources and then
co-added, with the advantage over an image-plane analysis that it
leads to reduced uncertainties.

Finally, stacking and $P(D)$ can take a role in the handling of
systematics. For example, in the 1--5$\sigma$ regime, {\it P(D)} has a
strong signal in terms of the source count at least, and if nothing
else, {\it P(D)} source-count analyses can show just what and when
Eddington bias overwhelms the data, so that one can decide \textit{a
  posteriori} whether to count discrete sources down to 6$\sigma$ or
3.2$\sigma$ (or at whichever threshold is suitable). It has played its
part in the characterization of CLEAN bias in FIRST (see
e.g.~\citealt{chapter-stil}). Seymour (in prep.) has proposed stacking
independent interferometer pointings in order to obtain the primary
beam and total noise.

\section{The Road to SKA}

\noindent
There are a number of systematics in $P(D)$ and stacking experiments
that must be understood during the analysis of pathfinder data. Here
we give a summary of the dominant systematics relevant to the
sub-threshold regime about which we \textit{currently} know.

\begin{enumerate}

\item CLEAN/snapshot bias; this pernicious effect occurs where
  instantaneous \textit{uv} coverage is limited. This could therefore
  be a concern for wide-field snapshot surveys, but less of a problem
  for longer-integration targeted deep fields such as those envisaged
  for SKA1-MID.
  \cite{white2007} were able to correct for the bias with a simple
  scaling factor, \textit{appearing} to be linear even to relatively
  faint flux densities.

\item Resolution bias; \cite{stil2014} pointed out that a bias is
  introduced when the synthesized beam is undersampled. So for
  stacking SKA1 data, one may require higher-resolution images than
  allowed for by simply requiring the synthesized beam to be
  Nyquist-sampled. One must give due consideration to astrometry.

\item Instrumental calibration, e.g.~primary beam model, will
  certainly be an issue for SKA1; as a sky-based phenomenon,
  characterization of the primary beam is critical for noise-regime
  analyses, especially as one wants to go to larger areas and/or
  mosaics and for polarization. Another example is time-dependent
  system temperatures (corresponding to spatially-varying map noise)
  for e.g.~VLBI. Joint solution of sky and systematics is beginning to
  be investigated (Lochner et al., in prep.) but is not routine. See
  also Smirnov~et~al.~(in prep.).

\item Sidelobe confusion (as opposed to classical confusion),
  is additional noise introduced into an image via imperfect source
  deconvolution within the image (i.e.~by all sources below the source
  subtraction cut-off limit) and from the (asymmetric) array response
  to sources outside the imaged field-of-view. \cite{ghosts} give an
  example of artifacts buried in the thermal noise. For instruments
  with a large field-of-view this is particularly challenging and the
  instrumental artifacts are hard to separate from classical
  confusion. This is where an accurate $P(D)$ and careful stacking
  analyses (section \ref{sec:stacking}) will couple to fully exploit
  the imaging of the SKA surveys.

\item Noise characterization is an issue for any survey; understanding
  of the noise structure is critical for obtaining the correct result
  in a $P(D)$ or stacking experiment. Uncertainties on source
  intensities could vary by position in the map (e.g.~the mosaicking
  eggbox sensitivity pattern), by depth (if the confusion noise begins
  to contribute in a stacking experiment, or if the dirty synthesized
  beam begins to enter the stack), or by resolution (confusion noise
  again). If one has some idea of the functional form of the noise, it
  can be fitted simultaneously with the physics quantities; debiasing
  a single-point average is rather harder. In a fully Bayesian
  analysis the likelihood function should be appropriate for the
  distribution of the measurement uncertainties.

\item Source clustering and sample variance, which \cite{hey13} showed
  how to limit in the survey-design phase. This has also been tackled
  in the submillimetre by \cite{viero2013}.

\item Simulations are of crucial importance to all these analyses (for
estimating biases due to clustering to give but one example). For
radio-continuum studies we use the work of \citet{wil08}. Note that
accounting for the (possible) background excess found by ARCADE2
\citep{fix11} may require new sub-$\mu$Jy populations
\citep{sei11,ver11,ver14}. These in turn would require simulations
differing from those available via the \citeauthor{wil08}
prescription, not yet formulated as we move into the pathfinder phase.

\end{enumerate}

\noindent
The SKA pathfinders such as LOFAR, MWA, MeerKAT and ASKAP themselves
will clearly play a key role both in allowing us to update our
source-count constraints, as well as in improving statistical
techniques that in turn can be adopted for SKA1 analyses, to which we
now turn our attention.

\section{Considerations for SKA}

\noindent
The integrated version of the count \citep{vernstrom2014} provides
important source-surface-density data for design of SKA pathfinder
surveys and for development of the SKA. However, extrapolation to
other survey wavelengths via a single spectral index must be
restricted to a factor of 2 at most, to say 1.4 and 5.0\,GHz, and even
then over only a limited flux-density range.  Counts outside this
frequency range differ markedly, and these differences have been
modelled by radio-source population syntheses,
e.g.~\citet{jw99,wil08,mas11}.

\begin{SCfigure}
\centering
\includegraphics [trim=0 20 0 20,clip,width =0.65\linewidth]{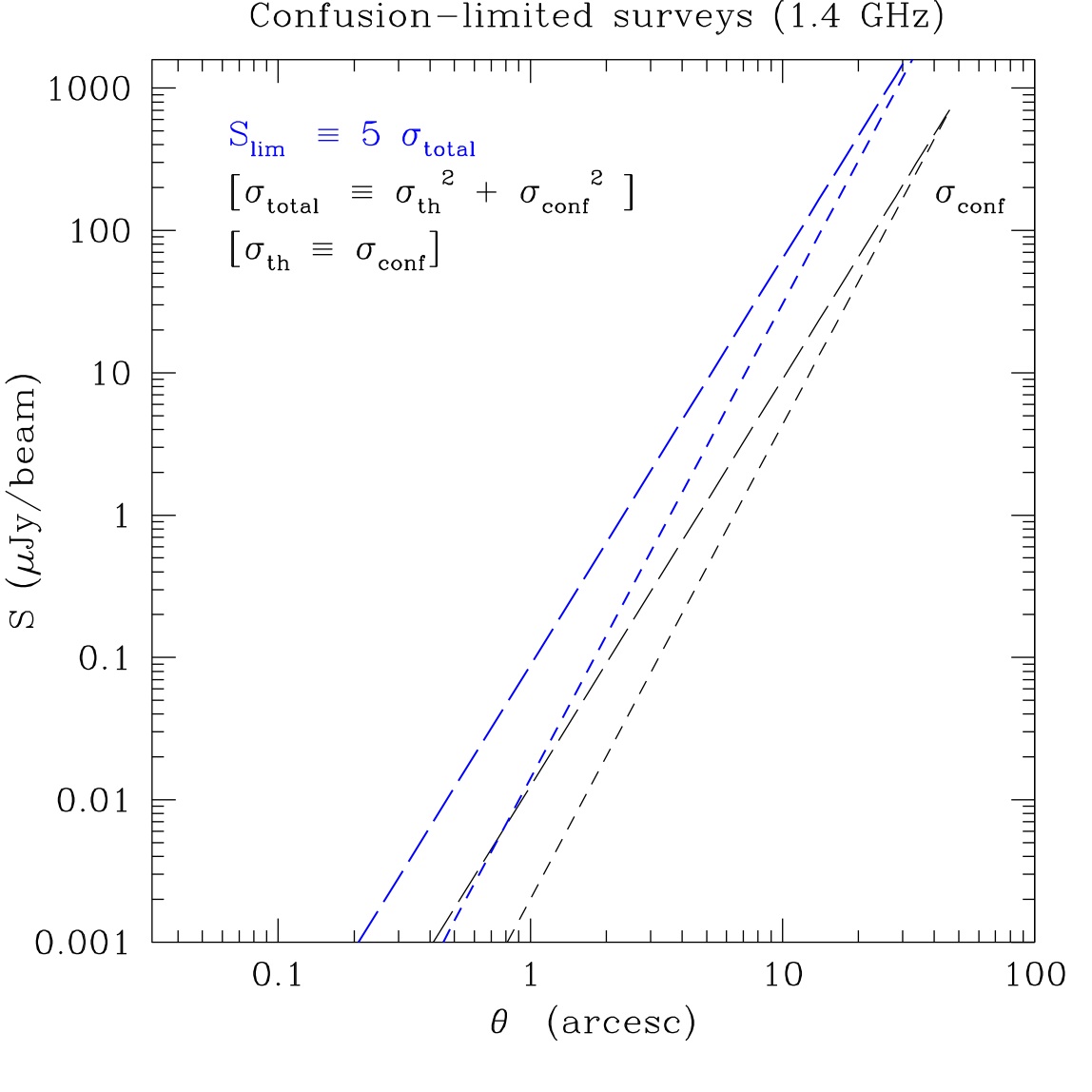}
\caption{1.4-GHz limits (dashed and dotted
  blue lines) of confusion-limited surveys, as a function of spatial
  resolution $\theta =$ HPBW. Two estimates of the confusion
  noise are considered. The first (dashed line) is the classical
  formulation, assuming the source counts derived by
  \citet{con12}, i.e.~$N(S)\approx 9000 S^{-1.7}$ Jy sr$^{-1}$ at 3
  GHz translating to $N(S)\approx 12000 S^{-1.7}$ Jy sr$^{-1}$ at 1.4
  GHz. The second (dotted line) uses the approximation
  of Eq.~27 of \citet{con12} assuming $\nu=1.4$\,GHz.}
\label{pran-sey}
\end{SCfigure}

\begin{figure}
\centering
\includegraphics [trim=0 000 0 000,clip,width =0.49\linewidth]{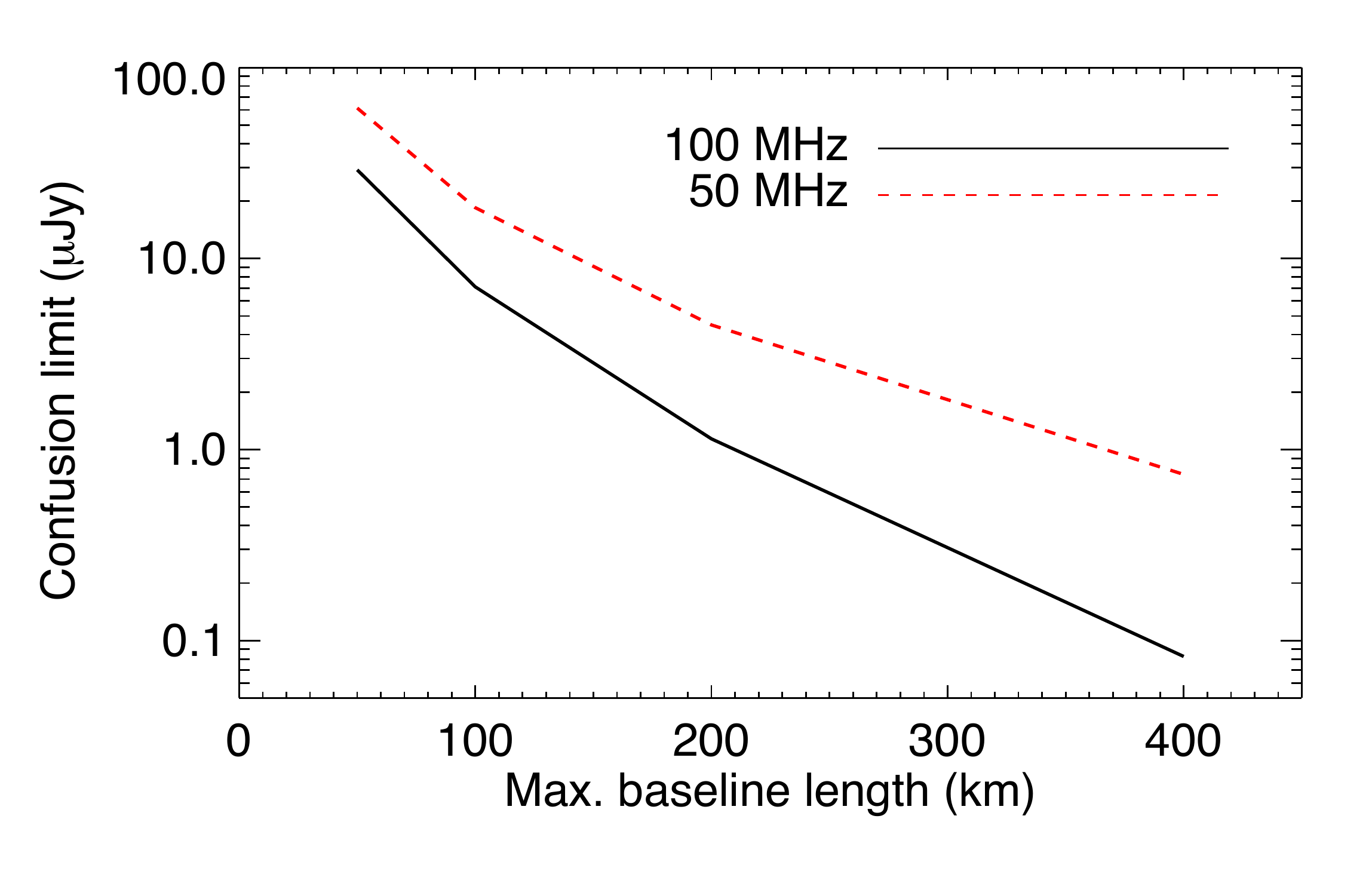}
\includegraphics [trim=0 000 0 000,clip,width =0.49\linewidth]{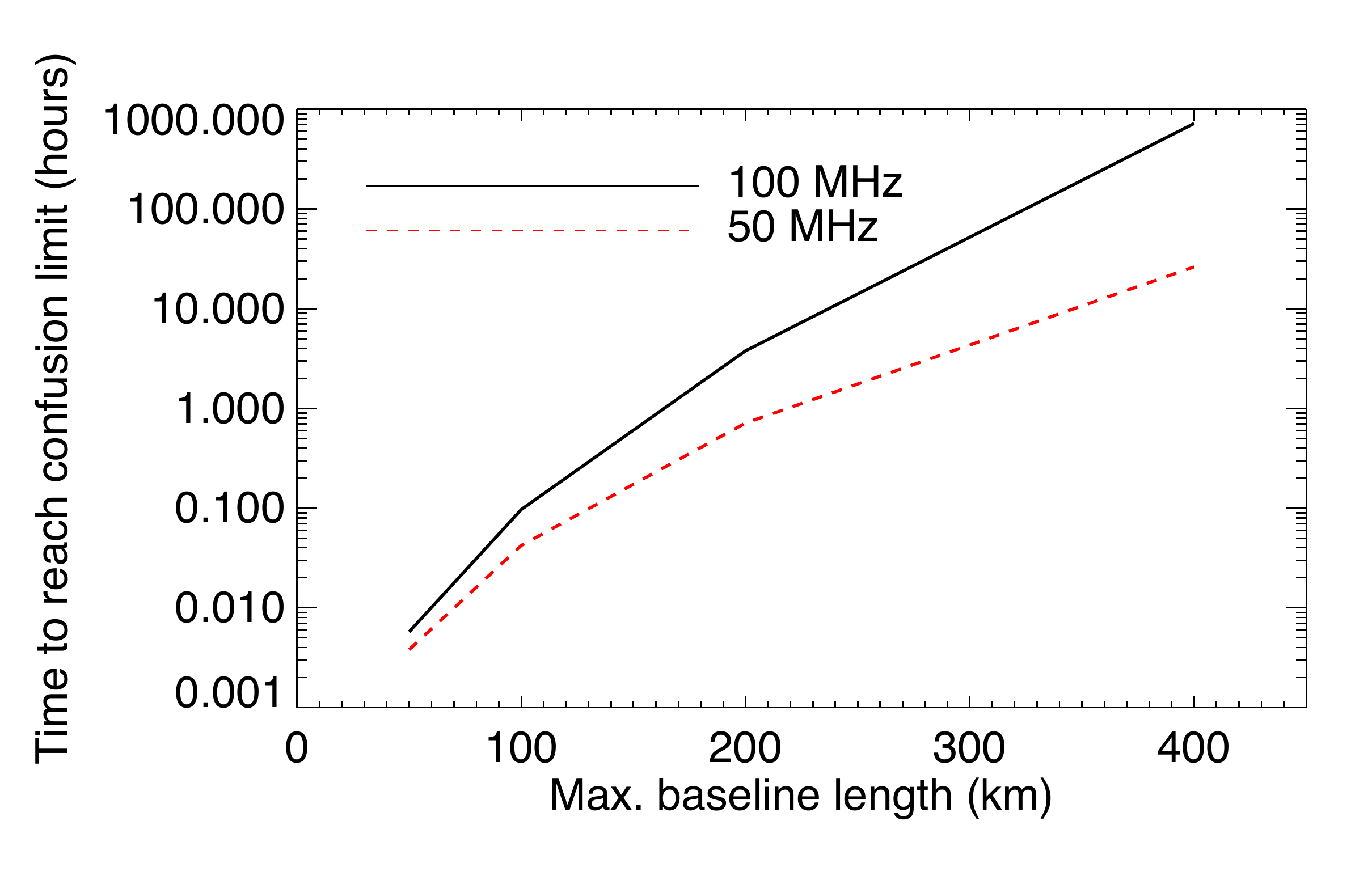}
\caption{{\it Left} -- the 1-$\sigma$ flux density at which the
  survey threshold/confusion limit (here taken as depth at which there is one
  $>3\sigma$ source per 20 beams) is reached for increasing maximum
  baseline. {\it Right} -- integration time per pointing needed to
  reach this limit for increasing maximum baseline. A maximum
  baseline of $>100$\,km is needed for surveys deeper than 10~$\mu$Jy.}
\label{em-huub}
\end{figure}

While it would be speculatory at this stage to predict the confusion
implications for the full SKA, various authors have made calculations
for SKA1, including present authors, \cite{chapter-summary}, and
\cite{condon-report}. Illustrative calculations are shown in
Figs~\ref{pran-sey} and \ref{em-huub} (for details of the reference
surveys refer to \citealt{chapter-summary}).  Although the details
remain to be calculated in the light of the newest deep count
estimates (e.g.~\citealt{vernstrom2014,ver14}), general statements can
be made. We deal with SKA1-LOW then SKA1-MID/SUR.

Firstly, $P(D)$ experiments at full sensitivity are only anticipated
for the lowest frequencies and possibly in the deepest fields, where
SKA1(-LOW) will see true confusion-limited data. The SKA will be
limited by dynamic range (see section \ref{sec:defns}), and
maximizing this will require care in configuration design
\citep{condon-report}. Second, SKA1-LOW will obtain its
confusion-limited data easily so that (a) survey design needs much
consideration and (b) confusion or $P(D)$ experiments will be
possible. These experiments may be of great significance, because of
the unknown contribution of very steep-spectrum populations favoured
at low frequencies. Such sources are hard to detect at faint
intensities/large distances of $\approx$1.4-GHz surveys because of
$K$-corrections. Their presence or otherwise in the low-frequency
source count will be telling in terms of their presumed
epoch-dependent luminosity function. Likewise there is likely to be
unique potential to probe star-forming galaxies to higher redshifts
than has previously been possible.

Estimates for the SKA1-LOW classical-confusion limit extrapolate
higher-frequency source counts (e.g.~1.4\,GHz) because of the paucity
of deep $\approx$150-MHz source-count data. While there is a steep
slope across the $\approx$100\,mJy--1\,Jy 150-MHz differential source
count, this will flatten if there is a sizeable, fainter source
population at $S_{150\,\mathrm{MHz}}<10\,$mJy. This behaviour is
predicted by the flattening of the 1.4-GHz differential source count
at $S_{1.4\,\mathrm{GHz}}\lesssim 2\,$mJy. Previous work, which has
adopted a standard spectral index to extrapolate from 1.4\,GHz to
150\,MHz in order to predict the low-frequency sky, could be
na\"{\i}ve: it assumes that the fainter population observed at
1.4\,GHz
obeys a standard spectral index ($\alpha=-0.7$) and/or that there is
no low frequency (only) very steep-spectrum faint population that is
undetectable at 1.4\,GHz.

For SKA1-MID, survey design needs careful consideration in terms of
whether confusion is to be avoided, or whether it is the object of
experiment (i.e.~$P(D)$). The detection thresholds of SKA1-MID/SUR
will be vastly lower than those of SKA1-LOW, mainly allowing for
stacking probes of radio populations selected in the formers'
higher-frequency all-sky surveys. But ultra-deep probes will demand an
adequate handling of source-clustering biases despite the high angular
resolutions the SKA will achieve at typical (GHz) survey wavelengths.

On the subject of panchromatic data, it is anticipated that stacking
will be undertaken in both wide and deep surveys, and Euclid and LSST
will almost certainly play a role here \citep{chapter-synergies}. If
millions of sources are detected by EUCLID, for example, over a wide
redshift range, and only (say) 20~per~cent are directly detected in
the radio, then stacking SKA1-MID data allows for the potential to
capture a substantial fraction of the galaxy population in order to
measure the luminosity functions of star-forming and AGN galaxies, and
as a function of environment (see elsewhere in this volume).

Characterization of any SKA1-MID-detected diffuse emission (such as
cluster halos and relics) that couples to shorter baselines will
necessarily require a good understanding of confusion noise (see
e.g.~\citealt{mcadam}). Stacking itself is highly applicable for
detecting such emission and constraining scaling relations
(e.g.~\citealt{act-stack}).

\bibliographystyle{apj}
\bibliography{noiseregime}

\end{document}